\documentclass[onecollarge,natbib]{svjour2}
\bibpunct{[}{]}{;}{n}{}{,} 
\smartqed  
\usepackage{graphicx}
\usepackage{amsmath}

\journalname{Few-Body Systems (EFB22)}
\begin{document}

\title{Mass-imbalanced Three-body Systems in 2D: bound states and the analytical approach to the adiabatic potential 
}


\author{F.F.~Bellotti \and
		T.~Frederico \and
		M.T.~Yamashita \and 
		D.V.~Fedorov \and
		A.S.~Jensen \and
		N.T.~Zinner
}


\institute{F.~F. Bellotti \at
              Instituto Tecnol\'{o}gico de Aeron\'autica, 12228-900, S\~ao Jos\'e dos Campos, SP, Brazil \\
              Department of Physics and Astronomy, Aarhus University, DK-8000 Aarhus C, Denmark \\
              Instituto de Fomento e Coordena\c{c}{\~a}o Industrial, 12228-901, S{\~a}o Jos{\'e} dos Campos, SP, Brazil \\
              Tel.: +55-12-39477147\\
              Fax: +55-12-39477111\\
              \email{ffbellotti@gmail.com}             \\
		\and
           T.~Frederico \at
              Instituto Tecnol\'{o}gico de Aeron\'autica, 12228-900, S\~ao Jos\'e dos Campos, SP, Brazil \\
        \and
           	M.T.~Yamashita  \at
    		  Instituto de F\'\i sica Te\'orica, UNESP - Univ Estadual Paulista, 01156-970, S\~ao Paulo, SP, Brazil \\
		\and
			D.V.~Fedorov, A.S.~Jensen, N.T.~Zinner \at
              Department of Physics and Astronomy, Aarhus University, DK-8000 Aarhus C, Denmark \\
}

\date{Received: date / Accepted: date}

\maketitle

\begin{abstract}
Three-body systems in two dimensions with zero-range interactions are considered for general masses and interaction strengths. The problem is formulated in momentum space and the numerical solution of the Schr\"odinger equation is used to study universal properties of such systems with respect to the bound-state energies. The number of universal bound states is represented in a form of boundaries in a mass-mass diagram. The number of bound states is strongly mass dependent and increases as one particle becomes much lighter than the other ones. This behavior is understood through an accurate analytical approximation to the adiabatic potential for one light particle and two heavy ones. 

\end{abstract}

\section{Introduction}
\label{intro}
The  properties of quantum systems change drastically with the number of space dimensions. For instance, the scattering length is not well defined for 2D systems \cite{adhikariAJoP1986} and the kinetic energy operator gives a negative (attractive) centrifugal barrier for 2D systems with zero total angular momentum, while the centrifugal barrier is always non-negative (zero or repulsive) for 3D systems. Another important difference between 2D and 3D systems is the occurrence of the Thomas collapse and the Efimov effect. These effects were predicted and observed for three identical bosons in 3D systems, but are absent in 2D. While three-identical bosons can have infinitely many bound states in 3D, only two three-body bound states are present in 2D \cite{adhikariPRA1988}.

Universal properties of mass-imbalanced three-body systems in 2D are studied using zero-range interactions in momentum space and only masses and energy ratios enter in the equations for bound states \cite{bellottiJoPB2011}. The critical values of these parameters (masses and two-body binding energies) allowing a given number of three-particle bound states with zero total angular momentum are determined in form of boundaries in a mass-mass diagram  (see Fig.~\ref{Graph01}) \cite{bellottiPRA2012}. The number of bound states is mass-dependent and increases as the mass of one of the particles decreases with respect to the other two \cite{bellottiJoPB2013}. This scenario is suitable handled in the Born-Oppenheimer (BO) approximation, which was already successfully implemented in 3D \cite{fonsecaNPA1979} and 2D \cite{limZfPAHaN1980}, aimed at the study of the Efimov effect in such systems. 

The theoretical prediction of a rich energy spectrum for highly mass-imbalanced systems combined with the rapid advance of the techniques in ultracold atomic experiments, which brings the possibility of trapping different atomic species together, asks for a deeper understanding of such systems. In view of that, the BO approximation for 2D three-body systems is re-visited under the mass-dependence perspective and the adiabatic potential is found to be mass-dependent, revealing an increasing number of bound states when the mass of one of the particles is decreased. This adiabatic potential is defined as the solution of a transcendental equation and an accurate analytic expression describing it is presented. The number of bound states for a heavy-heavy-light system is estimated as a function of the light-heavy mass ratio and infinitely many bound states are expected as this ratio approaches zero. However, for finite masses a finite number of bound states is found.

\section{Bound states}
\label{sec:1}
The 2D problem of three distinct particles of masses $m_a,m_b,m_c$, interacting with each other through an attractive zero-range interaction with two-body bound state energies $E_{ab},E_{ac},E_{bc}$ is considered. The $s-$wave three-body energy ($E_3$) is fully determined by these six parameters, namely two-body energies and masses. 

One advantage in the use of two-body energies instead of interaction strengths is that only mass and energy ratios enter in the equations. This means that the three-body energy divided by one of the two-body energies can be expressed as a function of four dimensionless parameters, i.e.
\begin{equation}
 \epsilon_3 = F_n\left(\frac{E_{b c}}{E_{a b}},
\frac{E_{a c}}{E_{a b}}, \frac{m_b}{m_a}, \frac{m_c}{m_a} \right)
 \equiv F_n\left(\epsilon_{b c},\epsilon_{a c},\frac{m_b}{m_a}, \frac{m_c}{m_a} \right),
\label{e40}
\end{equation}
where $\epsilon_3=E_3/E_{a b}$ is the scaled three-body energy and $\epsilon_{b c}=E_{b c}/E_{a b}$ and $\epsilon_{a c}=E_{a c}/E_{a b}$ are the scaled two-body energies. The universal functions, $F_n$, are labeled by the subscript $n$ to distinguish between ground, $n=0$, and excited states, $n>0$. 

This study is focused on two cases: $(i)$ for three pairs having the same interaction energy ($E_{a b}=E_{a c}=E_{b c}$) and,  $(ii)$ two pairs interacting with the same energy while the other one is not interacting ($E_{a b}=0\;,E_{a c}=E_{b c}$). Notice that in 2D, a non-interacting $ab$ system has $E_{a b}=0$. The understanding of a given system interacting with different energies is relevant, since pairwise interactions are tunable through Feshbach resonances. 

In both cases a diagram indicating the number of bound states as function of the masses of the particles in the three-body system is presented. Such diagrams can be useful tools to guide the search of three-body binding energies in 2D systems, since they show beforehand which systems have a richer energy spectrum.
\begin{figure*}
\centering
\includegraphics[width=0.3\textwidth,angle=-90]{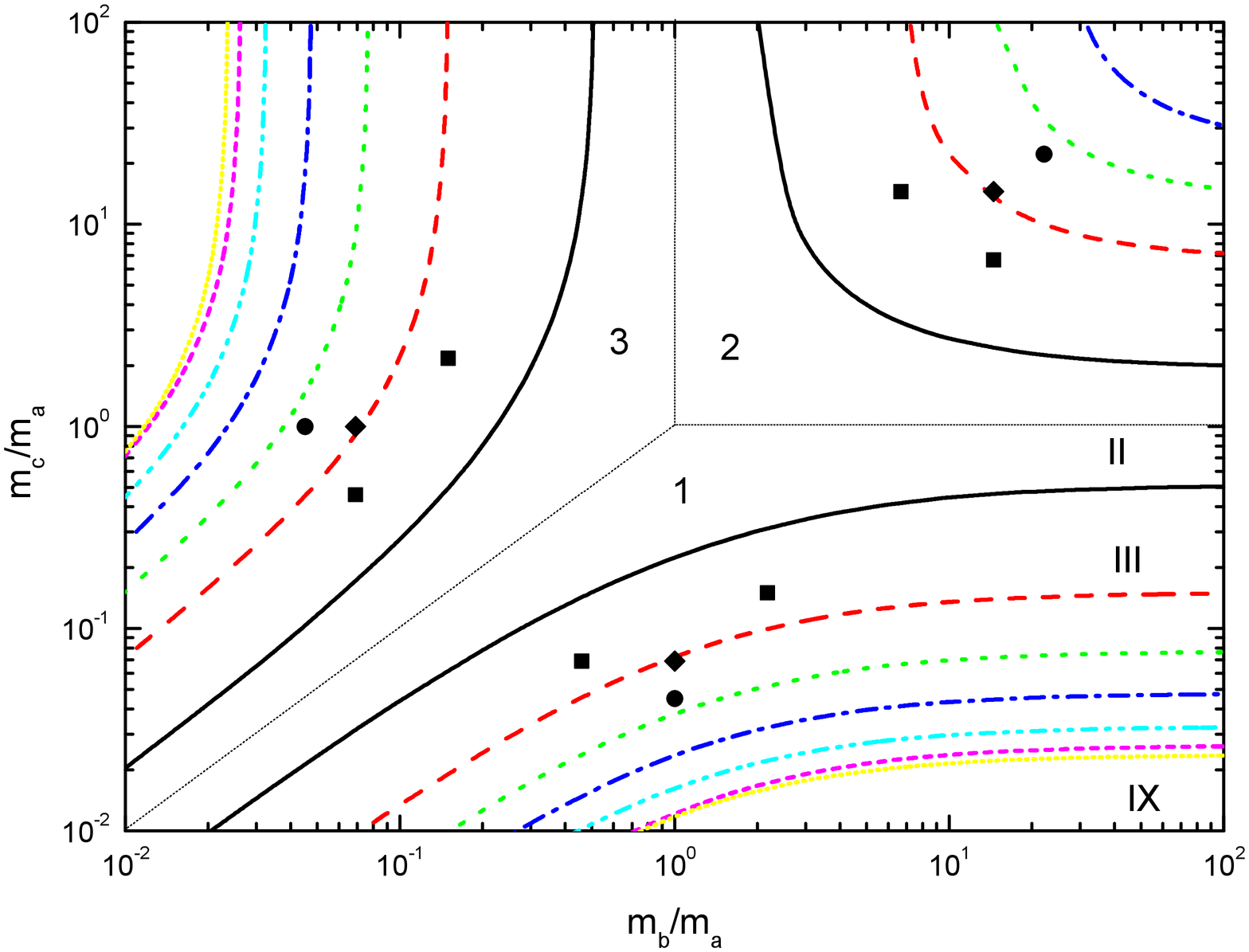}
\includegraphics[width=0.3\textwidth,angle=-90]{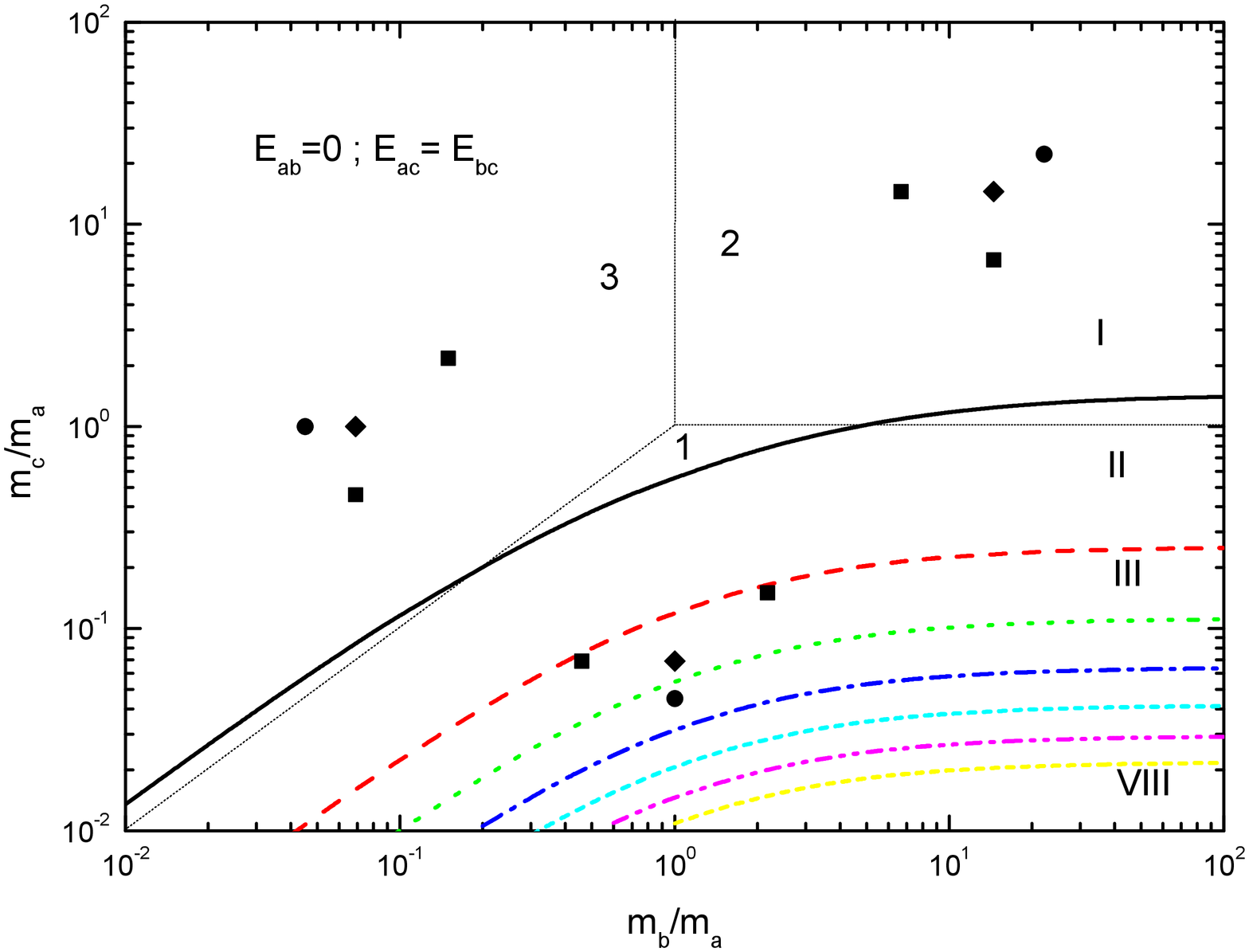}
\caption[dummy0]{Diagram of the number of three-body bound states as functions of two mass ratios, $\frac{m_b}{m_a}$ and $\frac{m_c}{m_a}$. The Roman numerals indicated the number of bound states in each region. The square ($^6$Li$^{40}$K$^{87}$Rb), circular ($^{6}$Li$^{133}$Cs$^{133}$Cs) and diamond ($^{6}$Li$^{87}$Rb$^{87}$Rb) points are the realistic cases  studied in \cite{bellottiPRA2012,bellottiJoPB2013}. The three sets of points are related by the symmetries in Eq.(~\ref{e40}). {\bf Left}: The three two-body energies are equal, $E_{ab}=E_{bc}=E_{ac}$. {\bf Right}: The two-body energies are  $E_{ab}=0,E_{ac}=E_{bc}=E_2$.}
\label{Graph01}
\end{figure*}

The richest energy spectrum is achieved, for any three-body system in 2D interacting with a contact interaction, when the three pairs have the same interaction energy. This diagram is showed in Fig.~\ref{Graph01}, where the central region ($m_a \approx m_b \approx m_c$) recovers the well-known result of two three-body bound states (indicated by II). Region II extends in three directions where two particles are heavier than the third one. These directions are indicated by dashed lines, which dived the graph into three symmetric regions. The three regions are obtained by permutation of the particles in Eq.~(\ref{e40}) and the symmetry is due to all the pairs having the same interaction energy.

Symmetry restricts the following analyses to just one region. Going downwards from the central point, regions with three (III), four (IV) etc. three-body bound states are achieved, indicating that the lighter one particle is compared to the other ones, the more bound states are available. Systems lying in region IV have a larger probability to have three-body bound states measured in cold atomic experiments than a system in region II. 

While the previous scenario ($E_{a b}=E_{a c}=E_{b c}$) is the most favorable to achieve a rich energy spectrum, in practice it can be very hard to set up this energy configuration in experiments. It was reported that mixtures of $^{6}$Li-$^{133}$Cs were successfully trapped in cold atomic experiments and three-body binding energies can be studied when the $^{133}$Cs-$^{133}$Cs is almost non-interacting \cite{reppPRA2013}. This configuration is described by choosing, for instance, $E_{ab}=0$ and keeping $E_{ac}=E_{bc}$. Region I, where only one three-body bound state is allowed, emerges in the middle of the figure, pushing the other lines away from the center. Excited states are only present in sector $1$, where the two non-interacting particles are heavier than the other one, i.e., $m_c<m_a$ and $m_c<m_b$, as illustrated in Fig.~\ref{Graph01}. 

The scenario of two heavy non-interacting particles and a light one also supports a rich energy spectrum and has the advantage to be accesible in laboratories in the near future. In this way, mixtures of $^{6}$Li-$^{133}$Cs and $^{6}$Li-$^{87}$Rb, which are of great interest in ongoing experiments, are good candidates for the study of 2D three-body systems.

\section{Adiabatic potential}
\label{sec:2}
It was shown in the previous section that a system composed of two heavy particles and a light one can have several energy levels. In this limit the interaction between the two heavy particles, which is described by an effective potential, is mediated by the light one. The Born-Oppenheimer (BO) approximation is used to calculate this potential. 

The BO approximation allows to split the three-body problem into two two-body problems: the dynamics of the heavy-heavy system and the motion of the light-particle with respect to the heavy-heavy center-of-mass. The adiabatic approximation says that it is possible to solve the dynamics of the light particle while the heavy ones are considered instantaneously at rest and in this way the eigenvalues of the light-particle equation plays the role of an effective potential for the heavy-heavy system. The effective potential is a function of masses and distance $\mathbf{R}$ between the heavy particles \cite{bellottiJoPB2013} and can be obtained from
\begin{eqnarray}
&\ln \left\vert \frac{ \epsilon(R)}{E_2 }\right\vert =2 K_0 \left(\sqrt{\frac{2m_{ab,c}|\epsilon(R)|}{\hbar^2}}R \right)  \ ,& \label{eq.51}
\end{eqnarray}
where $K_0$ is the modified Bessel function of second kind of order zero, $m_{ab,c}=m_c(m_a+m_b)/(m_a+m_b+m_c)$ and $E_2=E_{ac}=E_{bc}$.

The effective potential $\epsilon(R)$ is exactly defined as the solution of Eq.~(\ref{eq.51}), however a transcendental equation involving modified Bessel function is not intuitive at all. Two limiting expressions are found by expanding both sides of Eq.~(\ref{eq.51}) for small and large $R$. The result for $\left\vert  \epsilon_{asymp}(R)/E_2 \right\vert$ is
\begin{equation}
\frac{2 e^{-\gamma}}{s(R)} \;\;\; \text{for $R\to 0$}\;\;\;\;\;\; \text{and}\;\;\;\;\;\; 1+ \sqrt{2 \pi } \frac{e^{-s(R)}}{\sqrt{s(R)}}\;\;\; \text{for $R\to\infty$}, 
\label{adpot}  
\end{equation}
where $\gamma$ is Euler's constant and $s(R)=\sqrt{\frac{2m_{ab,c}\vert E_2\vert}{\hbar^2}} \ R$. 

For $R \to 0$ the effective potential resembles a hydrogen atom in 2D, where the pre-factor $1/\sqrt{m_c}$ makes the energy of the deepest states grow without boundaries when $m_c \to 0$. Furthermore, for $R \to \infty$, the potential is long-ranged and screened by a factor $\sqrt{m_c}$, which becomes less important for $m_c \to 0$. Therefore, an increasing number of bound states is expected when particle $c$ is much lighter than the other ones ($m_c \to 0$) since the adiabatic potential becomes more attractive and less screened in this limit. Still, these states will accumulate both at $R\to 0$, as the strength of the Coulomb-like potential increases, and at $R\to \infty$, where more states are allowed because the exponential moves to larger distances (Eq.(\ref{adpot})).  However, for finite $m_c$, still the number of bound states is finite.

\begin{figure}
\centering
\includegraphics[width=0.35\textwidth,angle=-90]{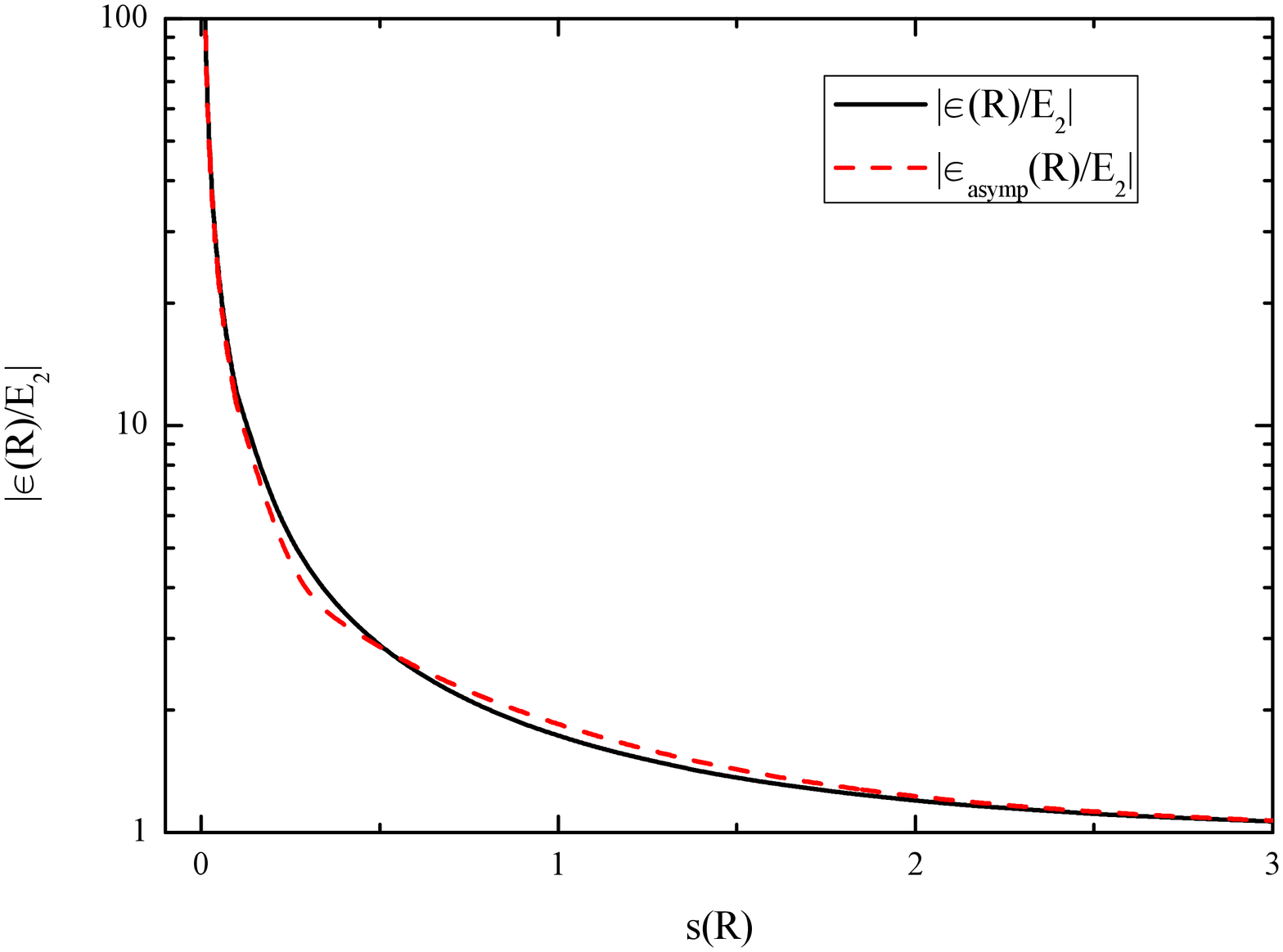}
\caption{ Exact numerical solution and adiabatic approximations for $\left\vert  \epsilon(R)/E_2 \right\vert$, respectively from Eqs.~\eqref{eq.51} and \eqref{adpot}, are shown as functions of the dimensionless coordinate $s(R)$. The limiting expression for $R \to 0$ in Eq.~\eqref{adpot} is plotted in the interval $0 < s(R) \leq 0.3$ and  the expression for $R \to \infty$ is plotted for $s(R) \geq 0.3$. } 
\label{Graph03}
\end{figure}

In spite of the fact that Eq.(\ref{adpot}) is valid in the extreme limits $R\to 0$ and $R\to \infty$, it perfectly reproduces the effective potential in almost all the range of the scaled coordinate $s(R)$, as shown in Fig.\ref{Graph03}. The adiabatic potential in Eq.(\ref{adpot}) can be treated in the JWKB approximation and the number of bound states ($N_B$) for a 2D system composed of two identical particles ($a$) and a distinct one ($b$) can be computed as function of the mass ratio $m=m_b/m_a$. For $m<1$ the result is $N_B=0.733/\sqrt{m}$. 

\begin{acknowledgements}
This work was partly supported by funds
provided by FAPESP (Funda\c c\~ao de Amparo \`a Pesquisa do Estado
de S\~ao Paulo) and CNPq (Conselho Nacional de Desenvolvimento
Cient\'\i fico e Tecnol\'ogico) of Brazil, and by the Danish 
Agency for Science, Technology, and Innovation.
\end{acknowledgements}

\end{document}